\documentclass[reprint, amsmath,amssymb, aps,superscriptaddress,prb, showpacs, showkeys]{revtex4-1}
\usepackage{graphicx}
\usepackage{dcolumn}
\usepackage{bm}
\usepackage[caption=false]{subfig}
\usepackage{color}
\definecolor{darkred}{rgb}{0.0, 0.0, 0.0}

\begin{document}
\title{Topological Invariant Tensor Renormalization Group Method for Edwards-Anderson Spin Glasses Model}

\author{Chuang Wang}
\email{wangchuang@itp.ac.cn}
\affiliation{State Key Laboratory of Statistical Physics, Institute of Theoretical
Physics, Chinese Academy of Sciences, Zhong-Guan-Cun East Road 55,
Beijing 100190, China}
\affiliation{University of Chinese Academy of Sciences, Beijing}
\author{Shao-Meng Qin}
\affiliation{State Key Laboratory of Statistical Physics, Institute of Theoretical
Physics, Chinese Academy of Sciences, Zhong-Guan-Cun East Road 55,
Beijing 100190, China}
\author{Hai-Jun Zhou}
\affiliation{State Key Laboratory of Statistical Physics, Institute of Theoretical
Physics, Chinese Academy of Sciences, Zhong-Guan-Cun East Road 55,
Beijing 100190, China}

\date{\today}

\begin{abstract}
Tensor renormalization group method (TRG) is a real space renormalization
group approach. It has been successfully applied to both classical
and quantum systems.  
In this paper, we study a disordered and frustrated system, the two-dimensional
Edward-Anderson model, by a new topological invariant TRG scheme.
We propose an approach to calculate the local magnetizations and nearest pair correlations
simultaneously. 
The Nishimori
multi-critical point predicted by the topological invariant TRG agrees
well with the recent Monte-Carlo results. 
The TRG schemes outperform the  mean field methods
on the calculation of the partition function.
We notice that it maybe obtain a negative partition function 
at sufficiently low temperatures.
However, the negative contribution can be neglected if the systems is large enough.
This topological invariant TRG can also be used to study 
three-dimensional spin glass systems.
\end{abstract}

\maketitle
\section{Introduction}
Exploring the Edwards-Anderson (EA) model \cite{Edwards1975}
is significant but extremely difficult. 
The nature of spin glasses on three-dimension is still a heat debate  
between the mean field picture and droplet picture
\cite{Parisi1983,Fisher1988,Yucesoy2012,Billoire,Thomas2011,ParisenToldin2011}.
over the past 30 years. 
For the two-dimensional (2D) model, besides its interests in statistical physics,
it has wild applications on
image processing \cite{Tanaka2003}, computer vision \cite{Derin1987,Sun2003}, 
which is usually referred to as the Markov random field \cite{Li1995} in the computer scientists community.
In this paper, we propose a coarse-graining method for EA model on 2D 
square lattice and calculate local physical quantities 
simultaneously by the tensor renormalization group (TRG) method.

TRG is a real space renormalization
group approach initially introduced by Levin and Nave \cite{Levin2007}
for classical ferromagnetic Ising spin systems on 2D regular lattices.
This method is an extension of the density matrix renormalization
group method for one-dimensional quantum systems \cite{Schollwock2005}.
The basic idea is to perform a coarse-graining process on a tensor
network. Matrix low rank approximation is used to cut the degree of
freedom of tensor indices up to a maximum value $D$ through the singular
value decomposition.

Shortly after the introduction of the initial TRG method, an improvement was
made by Xiang and co-authors \cite{Xie2009}, who proposed a backward
iteration to calculate the environment tensor and improved the results
by considering the effect of the environment. 
The TRG method has excellent performance
on the classical ferromagnetic Ising model, the Potts model \cite{Hinczewski2008}
and the diluted ferromagnetic model \cite{Guven2010,Guven2010a}, etc.
It also becomes a crucial tool to handle 2D quantum systems \cite{Jiang2008,Gu2008,Li2011}. 
Very recently a further
improvement, namely the topological invariant TRG method, was proposed
in the papers \cite{Xie2012, Garcia-Saez2013} to extend the TRG to
three-dimensional (3D) ferromagnetic Ising cases. 

Unlike the ferromagnetic Ising model, the EA spin glasses
model \cite{Edwards1975} is heterogeneous, disordered and frustrated. 
It is intrinsic hard.
The problems of finding a ground state of the
2D EA model with external field and the general 3D EA model are proved to belong to NP-hard class \cite{Barahona1982},
which is commonly believed that no algorithm can solve them within polynomial time.
In the previous study, 
the mean field approximation \cite{Yedidia2005,Zhou2011,Zhou2012,Wang2013,Lage-Castellanos2013} 
and Monte Carlo Sampling  \cite{Hasenbusch2008}, transfer matrix method \cite{Ozeki1987},
numerical exact algorithm for 2D without the external field \cite{Barahona1982,Thomas2013}
are used to calculate local properties for individual finite size instances. 
These methods are combined with finite size scaling 
to investigate the thermodynamics limit properties.
The duality relationship \cite{Nishimori2002,Ohzeki2009a,Ohzeki2011}
and real space renormalization methods \cite{Jorg2012} are also employed to study the phase diagram
and universality. 
TRG can be exploited in both of two roles. 
It can be served as an approximate calculator of physics quantities for a single instance,
and it may also be used as a new renormalization method to directly investigate critical phenomenon. 
We, here, focus on the former role.
To our best knowledges, there is no work on applying TRG on spin glasses until now. 

In this paper, we proposed two main approaches.
Firstly, we show a new topological invariant coarse-graining 
scheme based on the work \cite{Xie2012}. It avoids
two problems when the method \cite{Xie2012} is directly applied on EA model:
cutting extra freedom of indices
and inversing singular matrices.
In the ferromagnetic Ising model, these two problems don't exist. 
Secondly, we propose an approach to compute local physical quantities simultaneously.
For example, all single-spin magnetizations
can be calculated by a single sweep of coarse-graining procedure and backward procedure. 
These two approaches are also useful for other heterogeneous systems.
In the numerical calculation, TRG may get a negative value of the partition function at very low temperature, 
which is
the major difference between the spin glasses model 
and other heterogeneous systems \cite{Guven2010,Guven2010a}. 
We show that the contribution of the negative part is comparable to the error flunctuation for a large enough system, and therefore it can be neglected.  
In the high temperature region, TRG outperforms the mean field method, 
belief propagation and generalized belief propagation 
\cite{Yedidia2005,Zhou2011,Zhou2012,Wang2013,Lage-Castellanos2013}, while 
the mean field methods are failed in the lower temperature because of the convergence problems.
The Nishimori multi-critical point \cite{Nishimori1981,Nishimori2002,Takeda2005}
is calculated by our TRG scheme. 
The  results agree well with the
recent Monte-Carlo results \cite{Hasenbusch2008}. 
We emphasis that the original TRG method \cite{Levin2007} can also be directly applied to any 
heterogeneous systems, including EA model, similar to 
the works on diluted ferromagnetic model \cite{Guven2010,Guven2010a}. 
The advantage of topological invariant scheme is 
that it can be extended to 3D cases \cite{Xie2012, Garcia-Saez2013}.

The paper is arranged as follows. In the remainder of Section 1, we
introduce the EA model and show how to convert it to a tensor network.
In section 2, we demonstrate our topological invariant TRG procedure.
In section 3, we show how to calculate local physical quantities by
backward iteration. In section 4, we list some numerical results to
test the validation of this method. In section 5, we discuss 
the further improvement and  applications.

\section{The Models}
\subsection{The Edward-Anderson Model}

We consider the classical 2D EA model on a periodic square lattice
with discrete coupling constants. The system consists of N spins $\{\sigma_{i}\}$,
M coupling constants $\{J_{ij}\}$ and N local external fields $\{h_{i}\}$.
Each spin $\sigma_{i}$ takes value from $\{+1,-1\}$. The overall
spins state $\underline{\sigma}=(\sigma_{1},\sigma_{2},\ldots,\sigma_{N})$
is referred to as a configuration. The energy function is defined
as 
\begin{equation}
H(\underline{\sigma})=
-  \sum_{(ij)\in E}  J_{ij}  \sigma_{i}  \sigma_{j}  
-  \sum_{i\in V}  h_{i}  \sigma_{i}
\; ,
\end{equation}
where $E$ and $V$ denote the edge set and vertex set of the system, respectively.

For a single instance of the EA model, the coupling constants and
local external fields are fixed according to predefined distributions.
In this paper, the value of $J_{ij}$ is randomly chosen from the binomial distribution
$P(J_{ij})=p\delta(J_{ij},1)+(1-p)\delta(J_{ij},-1)$, where $\delta(x,y)$
is the Kronecker delta symbol, which is 1 if $x=y$, otherwise is $0$ . The model parameter $0.5\leq p\leq1$
alters the system ranging from the spin glass ($p=0.5$) to the pure
ferromagnetic system ($p=1$). The configuration $\underline{\sigma}$
is supposed to follow the Gibbs-Boltzmann distribution:
\[
p(\underline{\sigma}) = \frac{1}{Z} \exp 
\bigl[
  -\beta H(\underline{\sigma}) 
\bigr] 
\; ,
\]
where $Z=\sum_{\underline{\sigma}}\exp\left[-\beta H(\underline{\sigma})\right]$
is the partition function. It is useful to rewrite the distribution
as a production of a set of non-negative weight factors

\begin{equation}
  p(\underline{\sigma})
  =\frac{1}{Z}
  \prod_{(ij)\in E} \psi_{ij} (\sigma_{i},\sigma_{j})
  \prod_{i\in V} \psi_{i} (\sigma_{i})\ ,
\end{equation}
where the weight factors have the form 
$\psi_{ij}(\sigma_{i},\sigma_{j})=\exp[\beta J_{ij}\sigma_{i}\sigma_{j}]$,
$\psi_{i}(\sigma_{i})=\exp[\beta h_{i}\sigma_{i}]$. If all the external
random fields are zero, the partition function and the pair-spin correlations
can be calculated exactly in polynomial
time \cite{Kasteleyn1961,Barahona1982,Thomas2013}. However, for general
external fields $\{h_{i}\}$, the problem is proved to be in the
NP-hard class \cite{Barahona1982}.

\subsection{Tensor Networks}
\begin{figure}
  \subfloat[]{
   \includegraphics[scale=0.40]{tensor_network.eps}
   \label{tn1}
  }
 \subfloat[]{
   \includegraphics[scale=0.40]{tensor_network2.eps}
   \label{tn2}
 }
 \\
 \subfloat[]{
   \includegraphics[scale=0.40]{tensor_network3.eps}
   \label{tn3}
 }
  \caption{Construction of a tensor network: 
  (a) The neighborhood of a vertex $i$. 
  (b) Each matrix $\phi^{(ij)}$ is split into two matrices
  by the singular value decomposition, so that each vertex $i$ is now surrounded
  by four matrices which share the common index $s_{i}$. 
  (c) Summing over the index $s_{i}$, the neighbor four matrices 
  contract to be a tensor $T^{i}$.}
 \label{tn}
\end{figure}

Any two-body interaction system can be transformed into a tensor network, 
in which the partition function of the system is equal to the
trace of all the tensors. The transformation is not unique. Here we
show a symmetric method. The transformation of the EA model on 2D square
lattice at a site $i$ is illustrated in Fig. \ref{tn}. Firstly each
Ising spin $\sigma_{i}$ is mapped to a Boolean variable $s_{i}=(1-\sigma_{i})/2\in\{0,1\}$,
so that each weight factor $\psi_{ij}(\sigma_{i},\sigma_{j})$ can
be expressed as a matrix $\Phi^{(ij)}$, where the element in $s_{i}$-th
row and $s_{j}$-th column is $\Phi_{s_{i}s_{j}}^{(ij)}=\psi_{ij}(1-2s_{i},1-2s_{j})$.
Note that the C-programming-language convention is used, in which the
index starts from 0. Meanwhile, each external weight factor $\psi_{i}(\sigma_{i})$
of field $h_{i}$ is mapped to a vector $\Phi^{(i)}$, of which the
$s_{i}$-th element is $\Phi_{s_{i}}^{(i)}=\psi_{i}(1-2s_{i})$. Next
step, we perform the singular value decomposition on the matrix
$\Phi^{(ij)}$, such that
\begin{eqnarray}
  \Phi_{s_{i}s_{j}}^{(ij)} & = & 
  \sum_{s_{ij}}  U_{s_{i}s_{ij}}^{(ij)}  d_{s_{ij}}
  V_{s_{j}s_{ij}}^{(ij)} 
  \; ,
\end{eqnarray}
where the matrices $U^{(ij)}$, $V^{(ij)}$ are real orthogonal matrices
and the vector $\underline{d}= (d_0,d_1)$ stores singular values in descending order.
Each element in the vector $\underline{d}$ is non-negative. 
The new variable $s_{ij}\in\{0,1\}$ is the index of the singular vector
$\underline{d}$. 
Let $\tilde{U}_{s_{i}s_{ij}}^{(ij)}=U_{s_{i}s_{ij}}^{(ij)}d_{s_{ij}}^{\frac{1}{2}}$,
$\tilde{V}_{s_{j}s_{ij}}^{(ij)}=V_{s_{j}s_{ij}}^{(ij)}d_{s_{ij}}^{\frac{1}{2}}$.
Then we have
\[
\Phi_{s_{i}s_{j}}^{(ij)}=\sum_{s_{ij}}
\tilde{U}_{s_{i}s_{ij}}^{(ij)}\tilde{V}_{s_{j}s_{ij}}^{(ij)}\; .
\]
Now, each variable $i$ is surrounded by four matrices $\tilde{U}_{s_{i}s_{ij}}^{(ij)}$,
$\tilde{U}_{s_{i}s_{ik}}^{(ik)}$, $\tilde{V}_{s_{i}s_{il}}^{(il)}$,
$\tilde{V}_{s_{i}s_{im}}^{(i,m)}$, where $j$, $k$, $l$, $m$ are
labels of the neighbor spins of the spin $i$. Finally, we sum over
$s_{i}$ and get a tensor $T_{s_{ij}s_{ik}s_{il}s_{im}}^{i}$:
\begin{equation}
  T_{s_{ij}s_{ik}s_{il}s_{im}}^{i}=
  \sum_{s_{i}}  \tilde{U}_{s_{i}s_{ij}}^{(ij)}
  \tilde{U}_{s_{i}s_{ik}}^{(ik)}  \tilde{V}_{s_{i}s_{il}}^{(il)}
  \tilde{V}_{s_{i}s_{im}}^{(i,m)}  \Phi_{s_{i}}^{(i)} 
  \; .
\end{equation}
The partition function of the original system is equal to the result
obtained by tracing over all the indices of the tensors defined on lattice
sites:

\begin{equation}
  Z=\sum_{\{\underline{s}\}}\prod_{i}T_{s_{ij}s_{ik}s_{il}s_{im}}^{i} 
  \;.
  \label{eq:tn}
\end{equation}

We refer to the network of tensors constructed by the above procedure
as a tensor network. On the original
lattice, each vertex $i$ is associated with a tensor $T^{i}$, and each
edge $(ij)$ is associated with a tensor index $s_{ij}$. In graphical
language, the tensor network is similar to a factor graph model with
weight factors defined on vertices and state variables defined on
edges, but a key difference is that the elements of tensors
are not necessarily non-negative. In the following discussions,
we rewrite
the tensor indices as $i_{0},i_{1},i_{2},i_{3}$, i.e.,
$T_{i_{0}i_{1}i_{2}i_{3}}^{i}$
for notational simplicity.

\section{Tensor Coarse-Graining procedure}

There are several ways to implement the tensor 
coarse-graining procedure \cite{Levin2007,Xie2012,Garcia-Saez2013}.
Generally, each coarse-graining iteration consists of two steps. 
First is contracting two neighbor tensors into a new tensor with bigger indices freedom degree. 
It is an exact procedure. If there is no computation limitation, 
the exact partition function could be got by the iteration of these contractions.
Second is cutting the indices freedom degree approximately, 
so that the computation is tractable.

We introduce our method for 
the tensor network defined on a 2D square lattice with the periodic boundary
condition expressed as Eq.~\eqref{eq:tn}. 
At the first step,
each two vertical
neighbor-tensor pair $T,\; T^{\prime}$ are contracted 
as showed in Fig. \ref{ptrg_demo1}.
We sum over the common index $k$, 
and the pair $T,\; T^{\prime}$ is unified into one tensor $R$:
\begin{equation}
  R_{(i_{0},j_{0}),j_{1},(i_{2},j_{2}),i_{3}}=
  \sum_{k}T_{i_{0},k,i_{2},i_{3}}T_{j_{0},j_{1},j_{2},k}^{\prime} \; .
  \label{eq:Contract}
\end{equation}
The new tensor $R$ has 6 indices $i_{0}$, $i_{2}$, $i_{3}$, $j_{0}$,
$j_{1}$, $j_{2}$. We combine two indices in the same direction
$i_{0}$, $j_{0}$ as a union index $\hat{i}_{0}$ and $i_{2}$,
$j_{2}$ as another union index $\hat{i}_{2}$, 
 so that the number of indices of $R$ is
still $4$, i.e., 
$\hat{i}_{0},\ j_{1},\ \hat{i}_{2},\ i_{3}$. 
After the contraction, the topological structure of the square lattice is preserved,
and the y-direction length shrinks to half,
while the degrees of  
indices freedoms associated with the edges along the x-direction
increases to the square of the previous one. 

\begin{figure}
 \subfloat[]{
   \includegraphics[scale=0.5]{ptrg_demo_v2.eps} 
   \label{ptrg_demo1}
 }
 \subfloat[]{
   \includegraphics[scale=0.5]{ptrg_demo_2_v2.eps} 
   \label{ptrg_demo2}
 }
 \subfloat[]{
   \includegraphics[scale=0.5]{ptrg_demo_3_v2.eps} 
   \label{ptrg_demo3}
 }
\caption{\label{ptrg_demo}
(Color online) Demonstration of TRG: The top figure is the microscope of 
the circled region in the bottom figure. The two vertical tensors
$T$ and $T^{\prime}$ in (a) are contracted into one tensor $R$ in (b), 
and the associated two indices $i_0$ and $j_0$ of (a) are combined into
one index $\hat{i}_0$.  If the degree of freedoms of the index
 $\hat{i}_0$ is larger than the cut-off parameter $D$,
we use the singular value decomposition to truncate this index and obtain the approximate tensors
$\tilde{T}$ and $\tilde{T}^{\prime}$ in (c).
Bold lines indicates the freedom of associated indices are greater
then the others, when the freedom exceeds the cut-off parameter D.
}
\end{figure}

At the second step, the union
indices $\hat{i}_{0}$ and $\hat{i}_{2}$ will be truncated alternatively
along x-direction 
if their freedom degrees are greater then a given cut-off parameter $D$.
Specifically, let us 
consider the two horizontal neighbor tensors 
$R_{\hat{k},j_{1},\hat{i}_{2},i_{3}}$ and 
$R_{\hat{i}_{0}^{\prime},j_{1}^{\prime}, \hat{k},i_{3}^{\prime}}^{\prime}$
in Fig.~\ref{ptrg_demo2}, which share a same index $\hat{k}$.
We think  $R$ and $R^\prime$ as a sub-system in the tensor network
with the internal variable $\hat{k}$ and the boundary variables
$\{\hat{k},j_{1},\hat{i}_{2},i_{3}\} $ and 
$\{\hat{i}_{0}^{\prime},j_{1}^{\prime}, \hat{k},i_{3}^{\prime}\} $.
The boundary variables interact with  other tensors, 
which can be considered as the environment of the sub-system.
We are going to approximate the sub-system by another one 
with a fewer freedom degree of internal variable 
such that the interaction with environment is as similar as possible.
Mathematically, it is done by the lower rank matrix approximation.
We rearrange the indices order of the tensor $R$ as 
$j_{1},\hat{i}_{2},i_{3},\hat{k}$, and group the first three indices
as an unique index $\underline{i}=(j_{1},\hat{i}_{2},i_{3})$. 
Then tensor $R$ becomes a matrix ${R}_{\underline{i},\hat{k}}$. 
In the same way, we get the matrix 
${R}^\prime_{\hat{k},\underline{i}^\prime}$ from the tensor 
$R^\prime$, where 
$\underline{i}^\prime=(\hat{i}_{0}^{\prime},j_{1}^{\prime}, i_{3}^{\prime})$.
We sum over the common index $\hat{k}$ to get a new matrix ${A}$:
\begin{equation}
  {A}_{\underline{i},\underline{i}^\prime}
  =\sum_{\hat{k}} 
  {R}_{\underline{i},\hat{k}}
  {R}^\prime_{\hat{k},\underline{i}^\prime}
  \; .
\label{eq:mt_a}
\end{equation}
The sub-system is now expressed by the matrix $A$.
To exactly represent the boundary interaction, 
the minimum freedom degree of the internal variable is the rank of $A$.
A lower rank approximation is made by the singular value decomposition.
The matrix $A$ is decomposed in the reduced form by
\begin{equation}
  A_{\underline{i},\underline{i}^\prime} 
 =
 \sum_{k'=0}^{\text{rank}(A)-1}
 U_{\underline{i},k'}d_{k'}
 V_{\underline{i}^\prime,k'} \; .
 \label{eq:mt_a_svd}
\end{equation}
The reduced singular value decomposition discards the zero elements of the singular vector $d$, which has no contribution to the sub-system. 
In the numerical computation, singular values less than the criterion 
$d_i<\epsilon=10^{-12}$ are considered to be zero.
If the rank of $A$  is  greater than the cut-off parameter $D$, we only keep 
the largest $D$ singular values. Let $a^\prime=\min\{\text{rank}(A),D\}$. 
The approximation of $A$ is expressed as
\begin{equation}
  A_{\underline{i},\underline{i}^\prime}\approx 
  \tilde{A}_{\underline{i},\underline{i}^\prime} 
  =\sum_{k^\prime=0}^{a^\prime-1}
  \tilde{T}_{\underline{i},k^{\prime}}
  \tilde{T}_{\underline{i}^\prime,k^{\prime}}^{\prime}
  \; ,
  \label{eq:mt_a_svd2}
\end{equation}
where $\tilde{T}_{\underline{i},k^{\prime}}=
U_{\underline{i}^\prime,k'}d_{k'}^{\frac{1}{2}}$
and
$\tilde{T}_{\underline{i}^\prime,k^{\prime}}^\prime
=V_{\underline{i}^\prime,k'}d_{k}^{\frac{1}{2}}$.
The matrices $\tilde{T}$ and $\tilde{T}^\prime$ are non-singular,
and therefore their inverse matrices always exists.
This property will be used in the next section. 
Next, we expand the grouped indices $\underline{i}$ and $\underline{i}^\prime$
and rearrange the order of indices to recover the tensor 
$\tilde{T}_{\hat{k},j_{1},\hat{i}_{2},i_{3}}$ and 
$\tilde{T}_{\hat{i}_{0}^{\prime},j_{1}^{\prime}, \hat{k},i_{3}^{\prime}}^{\prime}$.

By this way, the tensors $R$ and $R^{\prime}$ in Fig.~\ref{ptrg_demo2}
are replaced by $\tilde{T}$ and $\tilde{T}^{\prime}$ in Fig.~\ref{ptrg_demo3},
and the common index $\hat{k}$ is replaced by $k'$ whose degree of freedom is no greater than $D$.
The above procedure of cutting off variable $\hat{k}$ by the singular value decomposition
guarantees that the matrix $\tilde{A}$ is a best approximation
of $A$ among all matrices with the rank no greater than $D$, if
the measure of the error is the Frobenius norm $\| A-\tilde{A}\| _F$. 
Note that $A$ is
a $D^{6}\times D^{6}$ matrix. The complexity of directly decomposition
of $A$ is $O(D^{18})$ . Considering that the rank of $A$ is at
most $D^{2}$, we could reduce the complexity into $O(D^{8})$. Details
are illustrated in the appendix A.

We now rotate the present tensor network $90^\circ$ in Fig.~\ref{ptrg_demo3}, 
and then it has the same local structure of the tensor network as the one at the first step 
in Fig.~\ref{ptrg_demo1}, while the length along x direction is reduced by half.
We repeat the step 1 and step 2 once more. 
The size of the tensor network shrinks half both in x and y directions.

This is the complete step of a coarse-graining procedure. We repeat it, 
until the tensor network is reduced small enough to be tractable by brute-force
summation to get the partition function. 
In this paper, the final size is $2\times2$.  

In practical, the value of elements of the tensors increase exponentially
during the TRG procedure. So we need scale the tensor after each
step. 
The scaling is forcing the maximum singular value 
of each $\tilde{A}^{(i)}$  in  Eq.~\eqref{eq:mt_a_svd2}
in the present layer tensor network to be a fixed
value $S_{m}$, and we save the logarithm of the scale factor
for the $i$'th matrix at the $l$'th step as
\begin{equation}
  \phi_l^{(i)}=\ln (d_0^{(i)}) -\ln (S_m),
\end{equation}
where $d_0^{(i)}$ is the maximum singular value of matrix $\tilde{A}^{(i)}$.
The total free energy density is 
\begin{equation}
  f(\beta)= -\frac{1}{N\beta} \left( \sum_l \sum_i \phi_l^{(i)}+\log Z_r \right),
  \label{eq:f_final}
\end{equation}
where $Z_r$ is the remaining scaled partition function calculated by contracting the final $2\times2$ tensor network.

{\color{darkred}
The cut-off parameter D controls the space of approximate tensors when performing the coarse-graining procedure.  
If D is infinite large, the coarse-graining process is exact. Generally larger D will  get more accurate results.
}
In terms of computational complexity, 
our topological invariant TRG scheme is 
of order $O(D^{8})$, while the original
method \cite{Levin2007} and the higher order TRG \cite{Xie2012} are $O(D^{6})$ and
$O(D^{7})$, respectively.
Practically, the precision in calculating the free energy is better
than the original one \cite{Levin2007} for the same cut-off parameter $D$. 
Our tensor coarse-graining method is based on the higher order TRG \cite{Xie2012},
where the exact contraction step is same, 
but the approximate truncation is different.
The higher order TRG method truncates all the indices associated
with x-direction edges by  the higher order singular value decomposition.
However, we found that such a truncation scheme can't
report a sufficiently precise free energy density value
for the EA spin glass model.
We only truncate the x-direction indices alternatively, 
while the remaining half of the x-direction indices will be
contracted in the next step, so they are not necessary to be truncated.

\section{Marginal probability and Backward Procedure}

The EA model has no translational symmetry, and therefore the
local magnetization depends on vertex position.
The marginal probability distribution
of a vertex $i$ is given by:
\begin{equation}
  P_{i}(s_{i})=\frac{1}{Z}\sum_{\text{all indices}}
  T^{i}(s_{i})\prod_{j\in V\setminus \{i\}}T^{j} \; ,
  \label{eq:marginal}
\end{equation}
where $s_i$ is related to the spin $\sigma_i$ by
$\sigma_i = 1- 2 s_i$ 
and the term, all indices, under the summation is referred to as all the indices
of every tensor in the tensor network $\{T^i|i\in V\}$,
and $T^{i}(s_{i})$ is a tensor at
vertex $i$ when its spin $\sigma_{i}$
is fixed to $1-2s_{i}$:
\begin{equation}
  T_{i_{0}i_{1}i_{2}i_{3}}^{i}(s_{i})
  =\tilde{U}_{s_{i}i_{0}}\tilde{U}_{s_{i}i_{2}}^{'}\tilde{V}_{s_{i}i_{3}}
  \tilde{V}_{s_{i}i_{1}}^{'}\Phi_{s_{i}}^{(i)} \; .
\end{equation}
As showed in Eq.~(\ref{eq:marginal}), $P_{i}(s_{i})$ can be computed
by ordinary TRG method for any $i$ in the tensor network with a
special tensor $T^{i}(s_{i})$. 
However, it is impractical to calculate
the marginal probabilities for all the vertices in this way. In this work
we use the backward iteration method  \cite{Xie2009} to
compute the marginal spin probability distribution functions for all
the vertices simultaneously.

We define the environment tensor,
or just called the environment, of a local tensor $T^{i}$ as
\begin{equation}
  M^{i}_{i_{0}i_{1}i_{2}i_{3}}=
  \sum_{ \substack{ \text{all indices}\\ 
    \text{except } i_{0},i_{1},i_{2},i_{3}}} 
  \prod_{j\in V\setminus i}T^{j}
  \; ,
\end{equation}
where the summation is taken over all the indices of the tensor network
expect the indices of the tensor $T^i$.
An environment $M^i$ has the same indices as its correspondent
tensor $T^i$. The partition function can be re-written as
\begin{equation}
  Z=\sum_{i_{0}i_{1}i_{2}i_{3}}T_{i_{0}i_{1}i_{2}i_{3}}^{i}
  M_{i_{0}i_{1}i_{2}i_{3}}^{i}\; .
  \label{eq:env}
\end{equation}
And the marginal probability distribution is expressed as
\begin{equation}
  P_{i}(s_{i})=
  \frac{1}{Z}
  \sum_{i_{0}i_{1}i_{2}i_{3}}
  T_{i_{0}i_{1}i_{2}i_{3}}^{i}(s_{i})
  M_{i_{0}i_{1}i_{2}i_{3}}^{i}\; .
  \label{eq:marginal-env}
\end{equation}
Similarly, the nearest-neighbor pair-wise marginal distribution $P_{ij}(s_{i},s_{j})$
can be also expressed as a summation between a pair of neighbor tensors
and the correspond environment:
\begin{eqnarray}
  P_{ij}(s_{i}, s_{j}) &= &\frac{1}{Z}
 \sum_{i_{0}j_{0}j_{1}i_{2}j_{2}i_{3}k}
  T_{i_{0},k,i_{2},i_{3}}^{i}(s_{i})
  T_{j_{0},j_{1},j_{2},k}^{j}(s_{j}) \nonumber \\
  & & \quad \times 
  \hat{M}_{(i_{0},j_{0}),j_{1},(i_{2},j_{2}),i_{3}}^{ij} \; ,
\end{eqnarray}
where $\hat{M}_{(i_{0},j_{0}),j_{1},(i_{2},j_{2}),i_{3}}^{ij}$ is
the environment of the tensor 
$R_{(i_{0},j_{0}),j_{1},(i_{2},j_{2}),i_{3}}
=\sum_{k}  T_{i_{0},k,i_{2},i_{3}}^{i}
T_{j_{0},j_{1},j_{2},k}^{j}$.

We calculate environments of a tensor network
at a more detailed level based on knowing the 
environments at a coarse-grained level,
which we called the backward iteration. 
We start from the final coarse-grained $2\times2$ tensor network
after finishing the forward TRG procedure. The corresponding environment
$M^{i}$ of a tensor $T^{i}$ at this level can be calculated directly
by tracing the other three tensors.

Given the environments 
$M^{\tilde{T}}$, $M^{\tilde{T}^\prime}$ of 
the tensors $\tilde{T}$, $\tilde{T}^\prime$ in Fig.~\ref{ptrg_demo3}, we 
now show how to calculate the environments $M^{T}$, $M^{T^\prime}$ of the
tensors $T$, $T^\prime$ at the detailed level in Fig.~\ref{ptrg_demo1}. 
The definition of indices is the same as described in the
previous section and shown in Fig.~\ref{ptrg_demo}. 
We start from the relation equation of the tensor $\tilde{T}$ and its environment
$M^{\tilde{T}}$ in Eq.~\eqref{eq:env}
\begin{align}
  Z  &=
  \sum_{k^{\prime},j_{1},\hat{i}_{2},i_{3}}
  \tilde{T}_{k^{\prime},j_{1},\hat{i}_{2},i_{3}}M_{k^{\prime},j_{1},\hat{i}_{2},i_{3}}
  \label{eq:en_tt}\\
   &=
   \sum_{k^{\prime},k^{\prime\prime},\underline{i}}
   \tilde{T}_{k^{\prime},\underline{i}}
  \delta_{k^{\prime\prime}}^{k^\prime}
  M_{k^{\prime\prime},\underline{i}}  \; ,
  \label{eq:en_tt2}
\end{align}
where in the second line we group the indices $(j_1,\hat{i}_2,i_3)$ 
as $\underline{i}$, 
and insert a Kronecker delta function  $\delta^{k^\prime}_{k^{\prime\prime}}$.

The tensor $\tilde{T}^{\prime}_{
      \hat{i}_{0}^{\prime},j_{1}^{\prime},k^{\prime},i_{3}^{\prime}
    }$ can be viewed as a matrix 
    $\tilde{T}^{\prime}_{k^{\prime}, \underline{i}^\prime}$
if we exchange the order of indices to
$k^{\prime},\hat{i}_{0}^{\prime},j_{1}^{\prime},i_{3}^{\prime}$ 
and group the indices 
$\hat{i}_{0}^{\prime},j_{1}^{\prime},i_{3}^{\prime}$ 
as $\underline{i}^\prime$. 
As mentioned in previous section, the matrix 
    $\tilde{T}^{\prime}_{k^{\prime}, \underline{i}^\prime}$
is always non-singular, we replace the Kronecker delta function 
in Eq.~\eqref{eq:en_tt2} by
\begin{equation}
  \delta_{k^{\prime\prime}}^{k^{\prime}}=
  \sum_{\underline{i}^\prime}
  \tilde{T}_{k^{\prime},\underline{i}^\prime}^\prime
  (\tilde{T}^{\prime-1})_{\underline{i}^\prime k^{\prime\prime}} \; ,
\end{equation}
where $\tilde{T}^{\prime-1}$ is the
inverse of the matrix $\tilde{T}^{\prime}$.
The partition function is then expressed by
\begin{equation}
  Z=\sum_{\underline{i},\underline{j}}
    A_{\underline{i}, \underline{i}^\prime}
    M^{\tilde{A}}_{\underline{i},\underline{i}^\prime}
  \; ,
\end{equation}
where $\tilde{A}_{\underline{i}, \underline{i}^\prime}=
    \sum_{k^{\prime}}
      \tilde{T}_{k^{\prime},\underline{i} }
      \tilde{T}^\prime_{k^{\prime},\underline{i}^\prime}
      $ 
is defined in Eq.~\eqref{eq:mt_a_svd2}, 
and $M^{\tilde{A}}$ is
the environment of $\tilde{A}$: 
\begin{equation}
  M^{\tilde{A}}_{\underline{i},\underline{i}^\prime}=
    \sum_{k^{\prime\prime}}
      \left(
       \tilde{T}^{\prime-1}
       \right)_{\underline{i}^\prime,k^{\prime\prime}}
       M_{k^{\prime\prime},\underline{i}}
\end{equation}
Since $\tilde{A}$ is the lower rank approximation of 
$A$, where $A_{\underline{i},\underline{i}^\prime}=\sum_{\hat{k}}
R_{\hat{k},\underline{i}}
R_{\hat{k},\underline{i}^\prime}^\prime
$ defined in Eq.~\eqref{eq:mt_a}, the environment $M^{\tilde{A}}$ is 
approximately the environment of $M^A$
\begin{equation}
  M^A\approx M^{\tilde{A}}
\end{equation}
The physical explanation of the above approximation is that,  
for a sub-system expressed by $\tilde{A}$  and its environment,
if we replace this sub-system with another sub-system $A$, 
which interacts with the environment in a very similar way 
with more internal variable states, 
the environment will not change too much.
From the relationship of $A$ and its environment, we get
\[
  Z=\sum_{\underline{i},\underline{i}^\prime,\hat{k}}
  R_{\hat{k},\underline{i}}
  R_{\hat{k},\underline{i}^\prime}^\prime
  M^A_{\underline{i}, \underline{i}^\prime} \; .
\]
The environments of $R$ and $R^\prime$ are obtained as
\begin{align}
  M^{R}_{\hat{k},\underline{i}}
    &=
    \sum_{\underline{i}^\prime}
    R_{\hat{k},\underline{i}^\prime}^\prime
    M^A_{\underline{i}, \underline{i}^\prime} \; ,
    \\
    M^{R^\prime}_{\hat{k},\underline{i}^\prime}
    &=
    \sum_{\underline{i}}
    R_{\hat{k},\underline{i}}
    M^A_{\underline{i}, \underline{i}^\prime} \; .
\end{align}

We expand the grouped indices $\underline{i}$, $\underline{i}^\prime$
of matrices $M^R$ and $M^{R^\prime}$
and exchange the indices to get the environments
$M^R_{\hat{k},j_{1},\hat{i}_{2},i_{3}}$ and 
$M^{R^\prime}_{\hat{i}_{0}^{\prime},j_{1}^{\prime}, \hat{k},i_{3}^{\prime}}$
of tensors $R$ and $R^\prime$.
This is the backward iteration of the cut-off step.

The backward iteration of the contraction step is more straightforward.
We unpack the indices $\hat{k}$ and
$\hat{i}_{2}$ of $R$ in the view before contraction, 
hence $R_{\hat{k},j_{1},\hat{i}_{2},i_{3}}\rightarrow R_{(i_{0},j_{0}),j_{1},(i_{2},j_{2}),i_{3}}
=\sum_{k}T_{i_{0},k,i_{2},i_{3}}T_{j_{0},j_{1},j_{2},k}^{\prime}$
as shown in Eq.~\eqref{eq:Contract}, in which the first index $\hat{i}_0$ is the
index $\hat{k}$ here. From the relation of the tensor $R$ and its environment
\[
Z=\sum_{i_{0} j_{0} j_{1} i_{2} j_{2} i_{3} k}
T_{i_{0},k,i_{2},i_{3}}T_{j_{0},j_{1},j_{2},k}^{\prime}
M_{(i_{0},j_{0}),j_{1},(i_{2},j_{2}),i_{3}}^{R} \; ,
\]
we can get the environments of $T$ and $T^{\prime}$ as
\begin{align}
  M_{i_{0},k,i_{2},i_{3}}^{T} & =
  \sum_{j_{1}j_{2}j_{3}}T_{j_{0},j_{1},j_{2},k}^{\prime}
  M_{(i_{0},j_{0}),j_{1},(i_{2},j_{2}),i_{3}}^{R}
  \; ,
  \\
  M_{j_{0},j_{1},j_{2},k}^{T^{\prime}} & =
  \sum_{i_{0}i_{1}i_{2}}T_{i_{0},k,i_{2},i_{3}}
  M_{(i_{0},j_{0}),j_{1},(i_{2},j_{2}),i_{3}}^{R}
  \; .
\end{align}

After the above two steps,
the environment matrix $M^{\hat{T}}$ is calculated by knowing
the environment matrix $M$ of higher coarse-grained level tensor
network. We repeat this process until 
the environment tensors of the original
tensor network are obtained. 
Then the marginal probability distributions can be calculated
from Eq.~(\ref{eq:marginal-env}). 
In practice,
we reduce the computational complexity by utilizing the fact
that the matrix $A$
is at most rank $D^{2}$.

The backward iteration is initially introduced 
to design a better  way
to do tensor coarse-graining by minimizing the change of the 
whole system with the environments \cite{Xie2009} 
on the ferromagnetic Ising model. 
This improvement can also be applied to EA model in the same way. 
We here exploit the backward iteration to calculate local physical quantities simultaneously.

\section{Numerical results }

\begin{figure}
  \includegraphics{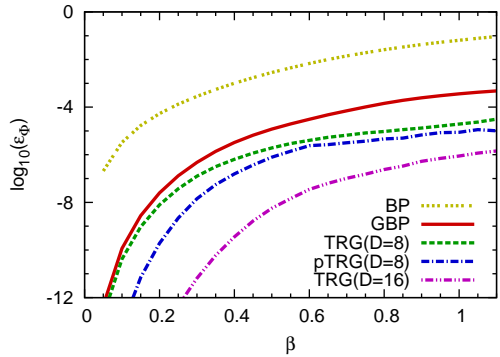}
  \caption{
    \label{comp_free_entropy}
    (Color online) 
    Comparison of the  error of $\frac{1}{N}\log Z$, calculated by
    our topological invariant TRG (pTRG) with $D=8$,
    the original TRG method (TRG) \cite{Levin2007} with $D=8,16$, and 
    the mean field approaches BP and GBP \cite{Wang2013,Lage-Castellanos2013}.
    The results are obtained by averaging,
    over 64 instances on a periodic square lattice with side length $L=64$.
  }
\end{figure}

We compared the partition function calculated by our topological invariant TRG 
with 
those obtained by the original TRG \cite{Levin2007} and mean field approach, 
belief propagation and generalized belief propagation (GBP)
\cite{Yedidia2005,Wang2013,Lage-Castellanos2013},
on the pure spin glass model without
external fields, i.e. $p=0.5$ and $h_i=0$. 
The exact partition function is calculated by the algorithm \cite{Barahona1982}.
The paramagnetic solutions of BP and GBP \cite{Wang2013,Lage-Castellanos2013} is included, 
which is the mean field method under 
the Bethe-Peierls approximation \cite{Bethe1935} and 
Kikuchi approximation \cite{Kikuchi1951} respectively. 
We measure the average error of the logarithm
partition function as 

\begin{equation}
 \epsilon_{\phi}=
 \frac{1}{N} \left\langle \left | \log(Z_{exact})-\log(Z) \right |  \right\rangle
\end{equation}
 over 64 instances with L=64
 in Fig. \ref{comp_free_entropy} in the region $\beta=1/T \in [0,1.1]$.
The results show that tensor renormalization approaches outperforms 
BP and GBP in several orders.
For the same cut-off parameter $D=8$, our topological invariant TRG are more accurate
than the original TRG. 
If one use a larger cut-off parameter
$D$, the results will be better, while the computation time will
increase dramatically.

\begin{figure}
  \subfloat[]{
   \includegraphics{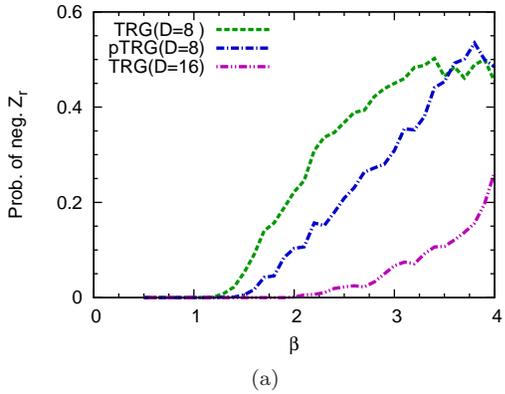}
   \label{failprob}
  }
  \\
  \subfloat[]{
    \includegraphics{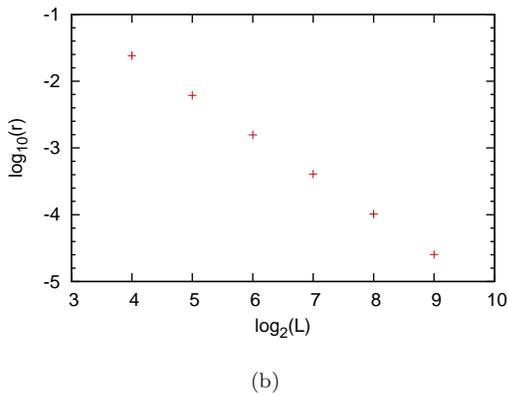}
    \label{phi_leading}
  }
  \caption{
    (Color online) 
    (a) The probability to obtain a negative $Z_r$  and
    (b) The ratio $r$ of $\log|Z_r|$  to the leading part at $\beta=1.5$.
  }
  \label{r_phi}
\end{figure}
At low temperatures $T$, i.e.  high inverse temperature $\beta=1/T$,
we found that the TRG procedures may
result in a negative partition function.
This phenomenon happens both in 
the original TRG \cite{Levin2007}
and our topological invariant TRG. 
We tested 128 instances with the inverse temperature $\beta$ ranging from
$0$ to $4.0$. 
The probability of negative partition function is 
are showed in Fig.~\ref{failprob}. 
A brief explanation is that the elements
in the tensors do not constrained to be non-negative and 
the lower rank matrix approximation makes
the final result to fluctuate around the exact partition function.
At low temperatures, the error is so large that
the scaled partition function $Z_r$ of the finial $2\times2$ tensor network 
turned out to be compatible with a negative value.
It seems a general limitation of TRG methods applying for the models
with frustrations. One could use larger cut-off parameter $D$ to reduce
the probability of negative results. 
If one only cares about the asymptotic
result for a large system, one could simply neglect the
negative part, since for infinite system the log partition function is dominated
by the scaling factors $\Phi_l^{(i)}$ in each forward iteration step rather than
the remaining contribution $Z_r$.
To clarify this point, we define the ratio of remaining log partition function 
and the leading part of the scaling factors as 
$r$,
\begin{equation}
  r=\left \langle \frac{ \log \left|Z_r \right|}{\sum_l \sum_i \Phi_l^{(i)}} \right \rangle,
  \label{eq:r}
\end{equation}
where $\langle \cdot \rangle$ means averaging over disorders. 
Numerically,  we averaged 128 instances.
As showed in Fig.~\ref{r_phi}, the contribution of remaining free entropy 
decreases as the system size increase almost linearly in the log-log scale.  
For a large system, it will be even
lower then the error, so that we can safely discard this term.
This phenomenon also indicates that we can investigate the EA model in the thermodynamic limit,
similar to the work on ferromagnetic Ising model  \cite{Hinczewski2008}.
Because of the heterogeneity, the properties of the system are captured
by infinite iterations of population of tensors rather than single tensor iteration. 
We leave the analysis of infinite systems in our future work.
\begin{figure}
  \includegraphics{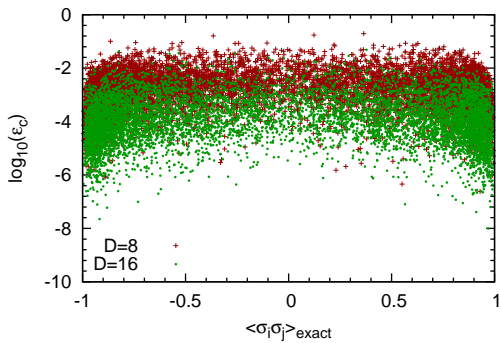}
  \caption{
  (Color online)
  Comparing the nearest neighbor correlations of a single instance L=64
  with exact results at $\beta=1$. The cut-off parameter is set to $D=8$
  and $16$.}
  \label{cor_xy}
\end{figure}

We plot all the nearest-pair-spin correlations of a typical single
instance compared by the numerical exact values which are calculated by numerical
differential of the free energy at $\beta=1.0$. 
The error is defined by
\begin{equation}
  \epsilon_c=\left( 
    \left|
      \langle \sigma_i \sigma_j \rangle_{\text{pTRG}}
      - \langle \sigma_i \sigma_j \rangle_{\text{exact}}
    \right|
  \right)
\end{equation}
Larger cut-off parameter $D$
will lead to better results, as shown in Fig.~\ref{cor_xy}. We do not
show the local magnetizations since they are always zero
because of the spin symmetry in the absence of no external fields.
\begin{figure}
  \includegraphics[]{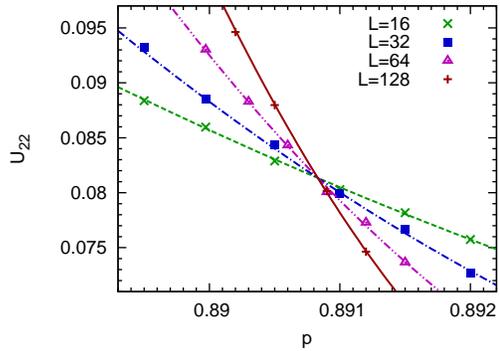}
\caption{(Color online) 
  \label{fig:mnp}
  Estimation MNP point by finite size scaling.
  {\color{darkred} 
  Lines are got by fitting Eq.~\eqref{eq:fit}. 
  }
  }
\end{figure}

The $p$-$T$ phase diagram of 2D EA model has been
extensively investigated in the papers 
\cite{Hasenbusch2008,Ozeki1987,Thomas2013,Nishimori1981,Nishimori2002,Takeda2005,Ohzeki2009a,Ohzeki2011,Jorg2012,Nobre2001,Bhatt1988}
and the references therein. 
There is no spin-glass phase occurred at finite temperature \cite{Bhatt1988}, 
while it undergo a para-ferro magnetic phase transition at low temperature $T$ and large $p$.
The system is in the paramagnetic phase when $0.5\leq p<p_c(T)$,  
and in  the ferromagnetic phase when $p_c(T)<p\leq 1$.
A special line $p_{\text{NL}}(T)=(\tanh(1/T)+1)/2$ is called 
the Nishimori line \cite{Nishimori1981}, on which some physical quantities 
can be calculated exactly. 
The multi-critical Nishimori point (MNP) is the crossing point of the Nishimori
line and the critical line $p_c(T)$.
We compute the MNP by locating the crossing point.

We use the topological invariant TRG as a tool to calculate magnetizations, 
and compute susceptibility $\chi$ by numerical differential 
{\color{darkred}
\begin{equation}
  \chi=\frac{d \sum_i \langle \sigma_i \rangle}{dh} \;,
\end{equation}
where $h$ is the external field and $\langle \cdot \rangle$ means averaging over the Boltzmann distribution, 
which can be quickly calculated by the marginal distribution Eq.~\eqref{eq:marginal-env} 
after the backward iteration.
}
The MNP point is estimated by finite size scaling stated in the work \cite{Hasenbusch2008}.
We measure the RG invariant quantity $U_{22}$, along Nishimori line near MNP, where
\begin{equation}
  U_{22}=
  {\color{darkred}
    \frac{ [ \chi^2 ]  }{ [ \chi ]^2}
  }
  -1\;,
\end{equation}
{\color{darkred}
where the square brackets are referred to as the average over the disorder, i.e. the couplings $\{J_{ij}\}$. We use $2\times 10^5$ instances for each point.
}
Then, the MNP point is got by fitting
\begin{equation}
  U_{22}=U_{22}^\ast + a_1 (p-p^\ast) L^{y_1} + a_2(p-p^\ast)^2 L^{2 y_1},
  \label{eq:fit}
\end{equation}
where $U_{22}^\ast$, $a_n$, $p^\ast$, $y_1$ are fitting parameters. 
{\color{darkred}
We fit the data with the lattice size $16\leq L\leq 128$ as showed in Fig.~\ref{fig:mnp}.
and estimiate the MNP point at $p^\ast=0.890830\pm 0.00022$, 
the exponent  $y_1 = 0.642\pm 0.022$
and other parameters $U_{22}=0.0813\pm 0.0003$, $a_1=-0.85 \pm 0.07 $, $a_2=6.5 \pm 2.6$. 
The chi-square test reports a small ratio of chi-square to the degree of freedom  $\chi^2 / DOF=7.2/17$, 
which show the fit model are good enough to describe the data.
We also test the fit by using different data group, for example $L\geq 32$, and $ L\leq 64$. 
All test are consistent with each other, excpet the data of $L=8$, which has strong finite size effect so that we discard it in all fit.
}
The susceptibilities $\chi$ are checked by using different differential steps 
$\delta h$ ranging from $10^{-6}$ to $10^{-3}$. 
For most of instances, they are insensitive to $\delta h$, and we set $\delta h=10^{-5}$.
While, a tiny fraction (about $10^{-4}$) depends on $\delta h$, and for these cases a larger $\delta h$ is used.
The location of MNP is not depend on the choices of $\delta h$.
Small portions of instances are also verified by averaging the two-point correlations. 
The comparison of the estimation MNP is showed in Table~\ref{tab:mnp}. 
The results agree well with the recent Monte Carlo method with finite size scaling \cite{Hasenbusch2008},
and the recent duality analysis inspired by hierarchical lattice \cite{Ohzeki2009a}.
We leave the discussion of re-entrance phenomena and strong disorder universality as the future work.
We emphasis that the role of TRG here is a new tool to calculate physical quantities. 
Compared other methods, the mean field estimation by BP and GBP on 2D EA model\cite{Wang2013, Lage-Castellanos2013}, 
it improve quite lot. 
\begin{table}
  \caption{\label{tab:mnp}Location of the multi-critical Nishimori point}
  \begin{ruledtabular}
    \begin{tabular}{ll}
        Methods & $p^\ast$ \\
         BP \cite{Lage-Castellanos2013} & 0.79  \\
         GBP \cite{Wang2013,Lage-Castellanos2013} & 0.85 \\
         Duality Analysis \cite{Nishimori2002}& 0.889972 \\
         Duality Analysis \cite{Ohzeki2009a}& 0.890813  \\
         pTRG & {\color{darkred}0.890830(22) } \\
         Monte-Carlo \cite{Hasenbusch2008} & {\color{darkred}0.89081(7) }\\
         {\color{darkred}Monte-Carlo \cite{ParisenToldin2009}} &{\color{darkred} 0.89083(3) }\\
    \end{tabular}
  \end{ruledtabular}
\end{table}

\section{Discussions and Conclusion }

In this paper, we applied the TRG on the 2D EA model, 
and proposed a novel topological invariant tensor coarse-graining procedure,
as well as an approach to calculate local physical quantities simultaneously.
Two problems hidden in the translation symmetric cases are solved.
We avoid to over-cut the freedom of indices in the coarse-graining procedure 
and avoid to inverse a singular matrix in backward iterations.
The backward iteration process was used to compute single spin marginal
probability distributions and nearest neighbor spin pair correlations.

We found that the TRG scheme is able to
compute the free energy and local correlations accurately if the
temperature is not very low. At low temperatures the TRG scheme might
lead to a negative value of the partition function.
We show that, for large systems, the main contribution of the partition function is the scaling factors during
the coarse-graining iteration, and
the negative remaining scaled partition function of the final $2\times2$ tensor networks can be discarded.
The successful estimation of the MNP location indicates TRG can be used in studying the critical phenomena
in a satisfied precision \cite{Hinczewski2008}, 
though originally TRG is considered only be applied to gapped phase\cite{Levin2007}. 
The present TRG scheme can't be applied to the case at zero temperature, 
because the SVD only preserves local optimal coarse-graining mode, 
and they are orthogonal in the further coarse-graining iteration and finally get zero partition function. 
It's an open question that whether TRG can be used at zero temperature problem. 
A further improvement can be made by 
considering the effect of environments, which is illustrated in the paper \cite{Xie2009,Xie2012} on
the ferromagnetic Ising model.  
{\color{darkred}
In principle, one can investigate the fixed point of TRG.
However, it may not get a good precision as we did in our paper,
because the advantage of TRG is its excellence performance on compute physical quantities rather than analyzing the fix point of the renormalization \cite{Efrati2014}.
}

The topic on the nature of spin glass phase on 3D
lattice is still rather active 
\cite{Parisi1983,Fisher1988,Yucesoy2012,Billoire,Thomas2011,ParisenToldin2011}.
The main method in most of the current studies 
is the Monte-Carlo Sampling. 
The topological invariant  coarse-graining iteration can be done in 3D cases 
by contracting tensors along one direction, 
and cut off the indices associated to the edges along the other two directions. 
Local physical quantities, for example the Edward-Anderson parameter, 
can be directly got as showed in this paper. 
The sample-to-sample overlap distribution or other non-local quantities would be 
estimated by TRG guided sampling, 
in the way that we fix the spins one by one according to its marginal probability. 
So, it presents an alternative way to investigate 3D spin glass models.

Another application is on
investigating combinatorial optimization problems
on finite-dimensional lattices or loopy random graphs.
TRG can be immediately applied on image segmentation and denoising \cite{Tanaka2003}. 
They share the same mathematical structure as 2D spin glasses model.
For random graph model, mean field method provided excellent solutions 
on mean-field like systems, such as 
local-tree like structured graph  \cite{Mezard2001} and fully connected graph \cite{Sherrington1975}. 
While, for the system rich in local loops, 
the mean field approximate may not quite accurate, for example small world networks, 
and many real networks. 
The extension of TRG on general graph provides a new insight and maybe 
another physics-contributed solution to such problems.
Similar to the belief propagation, 
the decimation \cite{Mezard2002} and reinforcement approaches  \cite{Chavas2005}
can be combined with TRG to get optimization solutions.

\begin{acknowledgments}
We thank Qiao-Ni Chen, Zhi-Yuan Xie, Jin-Fang Fan, 
Jack Raymond, Victor Martin-Mayor for helpful discussions, 
and thank Jack Raymond
for comments on an earlier version of the manuscript.
H.-J. Zhou were supported by the National Basic Research Program of China (No. 2013CB932804), the Knowledge Innovation Program of Chinese
Academy of Sciences (No.~KJCX2-EW-J02), and the National Science Foundation of China (grant Nos.~11121403, 11225526). The numerical simulations were
performed in the HPC computer cluster of ITP-CAS and we thank
Dr. Hongbo Jin for technical helps.
\end{acknowledgments}

\appendix
\section{Simplifying Singular Value Decomposition of Matrix $A$}

\begin{figure}
 \subfloat[]{\includegraphics{cutoff1.eps}

 }\quad{}\subfloat[]{\includegraphics{cutoff2.eps}

 }\quad{}\subfloat[]{\includegraphics{cutoff3.eps}

 }\quad{}\subfloat[]{\includegraphics{cutoff4.eps}

 }
\caption{Demonstration of simplifying cut-off step}
\end{figure}

We started from the definition of the matrix $A$ in Eq.~\eqref{eq:mt_a},
where $R$ and $R^{\prime}$ are tensors with four indices. We exchange
and combine the indices so that $R_{\hat{k},j_{1},\hat{i}_{2},i_{3}}$,
$R_{\hat{i}_{0}^{\prime},j_{1}^{\prime},\hat{k},i_{3}^{\prime}}^{\prime}$
change to be matrices $\hat{R}_{(j_{1},\hat{i}_{2},i_{3});\hat{k}}$,
$\hat{R}_{(\hat{i}_{0}^{\prime},j_{1}^{\prime},i_{3}^{\prime});\hat{k}}^{\prime}$.
For simplicity we write $\underline{i}=(j_{1},\hat{i}_{2},i_{3})$,
and $\underline{i}^{\prime}=(\hat{i}_{0}^{\prime},j_{1}^{\prime},i_{3}^{\prime})$.
Instead of multiplying $R$ and $R^{\prime}$, here we firstly decompose
them by the singular value decomposition
\begin{align}
  R_{\underline{i},k} & =\sum_{l}U_{\underline{i},l}d_{l}V_{l,k} 
  \; ,
  \\
  R_{k,\underline{i}^{\prime}}^{\prime} 
  & =\sum_{l^{\prime}}U_{k,l^{\prime}}^{\prime}
  d_{l}V_{l^{\prime},\underline{i}^{\prime}}^{\prime} 
  \; .
\end{align}
Let
\begin{equation}
  \tilde{A}_{l,l^{\prime}}=\sum_{k}d_{l}V_{l,k}
  U_{k,l^{\prime}}^{\prime}d_{l}
  \; .
\end{equation}
We decompose $\tilde{A}$ by the singular value decomposition
\begin{equation}
  \tilde{A}_{l,l^{\prime}}=\sum_{k^{'}}U_{l,k^{\prime}}^{A}
  d_{k^{\prime}}^{A}V_{l^{\prime},k^{\prime}}^{A}
  \; .
\end{equation}

Then tensors $\tilde{T}_{\underline{i},k'},\tilde{T}_{\underline{j},k'}^{\prime}$
in Eq~\eqref{eq:mt_a_svd2}, could be calculated by
\begin{align}
  \tilde{T}_{\underline{i},k'} & =\sum_{l}U_{\underline{i},l}
  U_{l,k^{\prime}}^{A}d_{l}^{A\frac{1}{2}}
  \; ,
  \\
  \tilde{T}_{\underline{i}^{\prime},k'}^{\prime} 
  & =\sum_{l^{\prime}}d_{l}^{A\frac{1}{2}}V_{l^{\prime},
  k^{\prime}}^{A}V_{l^{\prime},\underline{i}^{\prime}}^{\prime}
  \; .
\end{align}

The numerical SVD routines takes $O(mn^2)$ flops to decompose a $m\times n$ matrix ($m\geq n$) 
by Golub-Reinsch algorithm \cite{Golub1970}. 
The SVD routine in GNU Scientific Library is used in our numerical calculation.
The maximum size of matrix $\hat{R}$, and $\hat{R}^{\prime}$ are $D^4 \times D$. 
The SVD of these two matrices takes ${O(D^8)}$ flops, 
which take most computational complexity in the coarse-graining step,
while directly decomposing the $D^6 \times D^{6}$ matrix $A$ in Eq.~\eqref{eq:mt_a_svd} takes $O(D^{18})$ flops. 


%
\end{document}